\documentclass[submission, Phys]{SciPost}
\usepackage{amssymb}
\usepackage{amsthm}
\newtheorem{lemma}{Lemma}
\graphicspath{{pythonOGPaper/CorrFQ/}{oldVersion/}}

\begin{document}

\begin{center}{\Large \textbf{\color{scipostdeepblue}{\sf Symbolic operator growth and recursion method for spin-S Ising and q-state Potts models}}}\end{center}

\begin{center}
Igor Ermakov\textsuperscript{1,2,3,4$\star$}
\end{center}

\begin{center}

{\bf 1} Wilczek Quantum Center, Shanghai Institute for Advanced Studies, University of Science and Technology of China, Shanghai 201315, China
\\
{\bf 2} Hefei National Laboratory, Hefei 230088, China
\\
{\bf 3} Department of Mathematical Methods for Quantum Technologies, Steklov Mathematical Institute of Russian Academy of Sciences, 8 Gubkina St., Moscow 119991, Russia.
\\
{\bf 4} Russian Quantum Center, Skolkovo, Moscow 121205, Russia
\\[\baselineskip]
$\star$ \href{mailto:ermakov1054@yandex.ru}{\small ermakov1054@yandex.ru}
\end{center}

\section*{\centering Abstract}
\boldmath\textbf{
Operator growth is well understood for systems with a two-dimensional local Hilbert space, and much less so beyond them, where an operator can grow not only spatially but also in depth, inside the local algebra of each site. We compute the moments of infinite-temperature autocorrelation functions for the spin-$S$ Ising model and the $q$-state Potts model exactly and symbolically in the Hamiltonian parameters, in one, two and three dimensions. For the Ising model the moments are symbolic in the spin magnitude as well, so that a single computation covers arbitrary $S$. From them we obtain the Lanczos coefficients, rigorous Taylor bounds on the initial decay of the autocorrelation function, and its intermediate-time behavior via the recursion method. Our results support the Universal Operator Growth Hypothesis beyond local dimension two. Namely, the Lanczos coefficients grow linearly in the non-integrable cases and exhibit a clean square-root growth in the integrable one-dimensional Potts chain. For the spin-$S$ Ising model we show that every moment converges with growing spin to the corresponding moment of the classical spin model. The Lanczos coefficients therefore approach a limiting sequence with $1/S^2$ corrections, so that classicalization occurs at the level of the whole Lanczos sequence rather than of a single observable. This provides a justification for quasiclassical methods in infinite-temperature spin dynamics. All the results are exact, symbolic and obtained directly in the thermodynamic limit, and the computed moments are available in a public repository.
}\unboldmath

\vspace{10pt}
\noindent\rule{\textwidth}{1pt}
\tableofcontents\thispagestyle{fancy}
\noindent\rule{\textwidth}{1pt}
\vspace{10pt}

\section{Introduction}
The question of how an initially local quantum operator grows in many-body operator space has long attracted attention, both for its fundamental significance \cite{parker2019universal,zhang2019information,schuster2023operator,ermakov2025polynomially,dowling2025bridging,lunt2025emergent} and, more recently, for its growing role in computational physics \cite{ermakov2024unified,schuster2025polynomial,uskov2024quantum,de2024stochastic,nandy2025quantum,baiguera2026quantum,Loizeau2025,loizeau2025opening,shirokov2025quench,beguvsic2025real,angrisani2026simulating,rudolph2026pauli,ermakov2025symbolic,lychkovskiy2026qcommute}. Heisenberg-picture approaches based on operator dynamics are becoming a practical alternative to state-based methods, especially in regimes where tensor-network techniques \cite{verstraete2004matrix,zwolak2004mixed,tindall2024efficient} are fundamentally limited by volume-law entanglement. Recent developments have shown that approaches such as the recursion method \cite{uskov2024quantum,ermakov2025symbolic,shirokov2025quench,fullgraf2025lanczos} and operator propagation methods \cite{beguvsic2024fast,schuster2025polynomial,ermakov2024unified,de2024stochastic,miller2025simulation,rudolph2026pauli} can efficiently simulate certain classes of problems of quantum dynamics in two- \cite{uskov2024quantum,loizeau2025opening,teretenkov2024pseudomode} and even three-dimensional \cite{ermakov2025symbolic,beguvsic2025real,shirokov2025quench} systems.

Operator growth is well studied in systems with a two-dimensional local Hilbert space, such as interacting spin-$1/2$ systems~\cite{heveling2022numerically,schuster2023operator,de2024stochastic,uskov2024quantum} and fermions~\cite{ermakov2025symbolic}, where an operator can spread only spatially~\cite{swingle2018unscrambling,von2018operator,nahum2018operator,rakovszky2018diffusive,khemani2018operator,gopalakrishnan2018hydrodynamics}. Once the local dimension increases, the operator can also grow in depth, due to the larger local algebra on each site. Such cases have been examined far less, mostly for small systems outside the thermodynamic limit \cite{bhattacharjee2022krylov,bhattacharyya2023operator,kolganov2024streamlined}.

In this work we study operator growth in systems with a higher local Hilbert space dimension. In particular we consider the spin-$S$ Ising model and the $q$-state Potts model. The former has a natural classical limit at high spin $S\rightarrow\infty$, which makes it well suited for studying the quantum-to-classical transition at the level of operators. The latter has no well-defined classical limit and therefore calls for a different treatment. We address the technical aspects of this difference in detail below. Higher-spin systems are of interest both for fundamental questions such as quantum many-body scars and quantum-classical correspondence \cite{schecter2019weak,ermakov2025classical,hu2025krylov}, and for their role in qudit-based technologies~\cite{wang2020qudits,chi2022programmable}.

Our first motivation is to directly test the Universal Operator Growth Hypothesis (UOGH) \cite{Liu_1990_Infinite-temperature,Florencio_1992_Quantum,Zobov_2006_Second,Elsayed_2014_Signatures,Bouch_2015_Complex,parker2019universal} in systems with a local Hilbert space dimension larger than two. While the UOGH has been extensively tested in spin-$1/2$ models, its confirmation at higher local dimension has been missing, and here we fill this gap.

The second natural question is that of the classical limit. As the spin grows, quantities of a model with a well-defined classical limit are expected to approach their classical counterparts \cite{bolivar2004quantum,lichtenberg2013regular,kumari2018quantum,wang2021quantum}. Remarkably, we find that classicalization occurs at the level of operator growth as well. Namely, for the spin-$S$ Ising model we show that every moment has a finite limit as the spin grows, equal to the corresponding moment of the classical spin model, so that the Lanczos coefficients converge to a limiting sequence with $1/S^2$ corrections. At early times this convergence is fast. These results give a deeper justification for the performance of quasiclassical methods at $S>1/2$ \cite{polkovnikov2010phase,elsayed2013regression,schachenmayer2015many,zhu2019generalized,ermakov2025classical,starkov2020free,kempa2026random}.

Finally, operator growth can serve as an additional tool to distinguish integrable from non-integrable dynamics of a given observable \cite{bhattacharjee2022krylov,ermakov2025polynomially}. The $q$-state Potts model provides a clean example. It is integrable in one dimension~\cite{wu1982potts}, where we find sublinear growth of Lanczos coefficients, and non-integrable in two and three dimensions, where the growth becomes linear as predicted by the UOGH.

For the spin-$S$ Ising and $q$-state Potts models we compute symbolically the moments of infinite-temperature autocorrelation functions. From these moments we extract the Lanczos coefficients, the standard tool for studying operator growth~\cite{viswanath1994recursion,parker2019universal}. We also study the dynamics of the infinite-temperature autocorrelation function itself. For its early-time behavior we obtain rigorous benchmarks via Taylor bounds~\cite{Platz_1973_Rigorous,Roldan_1986_Dynamic,Brandt_1986_High,Bohm_1992_Dynamic}, for later times we employ the recursion method~\cite{Dupuis_1967_Moment,viswanath1994recursion}. All the results are obtained directly in the thermodynamic limit. The symbolic moments are exact, and the Lanczos coefficients follow from them in exact arithmetic, free of the numerical instabilities usual for the Lanczos algorithm.

A major technical challenge for studying operator growth lies in the complexity of nested commutator evaluations. To date, efficient tools exist for operator computations in the Pauli string \cite{Loizeau2025,lychkovskiy2026qcommute} and Majorana string \cite{ermakov2025symbolic,miller2025simulation} bases. Here we develop a symbolic engine that works in two bases suitable for higher spins. The first is the non-orthogonal basis of spin monomials, the second is the orthogonal basis of generalized Pauli strings. Each of them suits a different model. For the spin-$S$ Ising model we use spin monomials, whose algebra does not depend on $S$, and obtain results that are symbolic not only in the interaction parameters but in the spin magnitude itself. In other words, the resulting expressions are valid for arbitrary spin. For the Potts model the orthogonal basis of generalized Pauli strings is more convenient. Symbolic calculations for arbitrary spin and arbitrary Hamiltonian parameters have attracted interest for a long time \cite{van1948dipolar,rushbrooke1958curie,schmidt2011eighth,lohmann2014tenth}. Here we reach considerable depths, and we are not aware of comparable symbolic calculations.

The paper is organized as follows. Section~\ref{sec:background} collects the background: operator growth, the rigorous bounds on $C(t)$, the recursion method, and the representation of operators we use. Section~\ref{sec:basis} introduces the two bases of the spin operator space. Section~\ref{sec:ising} presents the results for the spin-$S$ Ising model, including the classicalization of the Lanczos coefficients, and Section~\ref{sec:potts} the results for the $q$-state Potts model. Section~\ref{sec:discussion} contains the discussion. Appendix~\ref{app:traces} describes the exact evaluation of scalar products in the non-orthogonal basis, and Appendix~\ref{app:thermlim} explains how the moments of the extensive observable are obtained from a single local operator directly in the thermodynamic limit. All the computed moments are available in a public repository~\cite{symboliclab}.

\section{Background}
\label{sec:background}

In this section we summarize the elements of operator growth theory used throughout the paper. We first recall the relation between nested commutators, autocorrelation moments, and Lanczos coefficients, then discuss the recursion method and the operator representations used in symbolic calculations.

\subsection{Operator growth}

We are interested in the Heisenberg evolution of an initially local operator $A$. In the Heisenberg picture,
$A(t)=e^{itH}Ae^{-itH}$. Introducing the Liouvillian superoperator $\mathcal{L}\equiv[H,\cdot]$, the equation of motion becomes $\partial_t A(t)=i\mathcal{L}A(t)$, with the formal solution $A(t)=e^{it\mathcal{L}}A$. Equivalently,
\begin{align}
    \label{nestedForm}
    A(t)=\sum\limits^\infty_{m=0}\frac{(it)^m}{m!}[H,A]^{(m)},
\end{align}
where $[H,A]^{(m)}\equiv\mathcal{L}^mA$ is the $m$-fold, or nested, commutator. This expansion is the starting point for operator-growth methods and exposes their main computational bottleneck: the repeated evaluation of commutators in an exponentially large many-body operator space.

We use the infinite-temperature scalar product
\begin{align}
    \label{scalProd}
    (A|B)\equiv\text{tr}(A^\dagger B)/\text{dim}(\mathcal{H}),
\end{align}
where $\text{dim}(\mathcal{H})$ is the finite Hilbert-space dimension. This scalar product induces the norm $\|A\|=\sqrt{(A|A)}$. The corresponding infinite-temperature autocorrelation function is
\begin{align}
    \label{autocorrF}
    C(t)\equiv(A(t)|A)/\lVert A\rVert^2.
\end{align}
It measures the overlap between the evolved operator and the initial operator. Taylor-expanding in $t$ and using Eq.~(\ref{nestedForm}) gives
\begin{align}
    \label{corrFtaylor}
    C(t)=\sum\limits^\infty_{m=0}(-1)^m\frac{\mu_{2m}}{(2m)!}t^{2m},
\end{align}
where
\begin{align}
    \label{moms}
    \mu_{2m}\equiv\frac{(\mathcal{L}^{2m}A|A)}{\lVert A\rVert^2}=\frac{(\mathcal{L}^{m}A|\mathcal{L}^{m}A)}{\lVert A\rVert^2}=\frac{\lVert [H,A]^{(m)}\rVert^2}{\lVert A\rVert^2},
\end{align}
are the even moments of $C(t)$. Odd moments vanish for the Hermitian observables considered here.

Another standard approach to operator growth uses the orthogonal Lanczos basis $\{A_n\}$, $n=0,1,2,\dots$, defined iteratively by $|A_0)=\|A\|^{-1}|A)$, $|A_1)=\mathcal{L}|A_0)$, and
\begin{align}
    \label{lancz}
    &b_n=\|A_n\|, \qquad n=0,1,2,\dots, \\
    &|A_n)=b^{-1}_{n-1}\mathcal{L}|A_{n-1})-b_{n-1}b^{-1}_{n-2}|A_{n-2}), \quad n=2,3,\dots\nonumber
\end{align}
The coefficients $b_n$, known as the Lanczos coefficients, encode detailed information about the operator dynamics. They are related to the moments~(\ref{moms}) through Hankel determinants~\cite{sanchez1978spectrum}, and to $C(t)$ through the coupled equations
\begin{align}
    \label{diffEqCt}
    &\partial_t\varphi_n(t)=-b_{n+1}\varphi_{n+1}(t)+b_n\varphi_{n-1}(t),\quad n=0,1,2,\dots\nonumber \\
    &C(t)=\varphi_0(t),
\end{align}
where $\varphi_{-1}(t)\equiv 0$ and $\varphi_n(0)=\delta_{0n}$.

Thus the autocorrelation function $C(t)$, its moments $\mu_{2m}$, and the Lanczos coefficients $b_n$ are equivalent ways of characterizing the growth of $A$.

\subsection{Bounds on the correlation function}
\label{sec:intro_bounds}

The first $n_\text{max}$ moments give rigorous bounds on $C(t)$. In particular, it is a well-known result \cite{Platz_1973_Rigorous,Roldan_1986_Dynamic,Brandt_1986_High,Bohm_1992_Dynamic,uskov2024quantum} that
\begin{align}\label{bounds}
    P_{4l\pm 2}(t)\leq C(t)\leq P_{4l}(t),
\end{align}
where $P_m(t)$ is the Taylor polynomial of $C(t)$ of degree $m$, built from the moments up to order $m$. Consecutive even-order truncations therefore bracket $C(t)$ from below and above. For infinite-temperature autocorrelation functions these bounds are tight at short times and loosen as $t$ grows. The Taylor polynomials diverge beyond the convergence radius $t^\ast$, which is finite in two and three dimensions~\cite{parker2019universal}.

Even within this finite radius, the bounds give $C(t)$ directly in the thermodynamic limit, in regimes far beyond the reach of exact diagonalization, especially in two and three dimensions. This follows from the local character of operator growth: each moment $\mu_{2m}$ involves only a finite region of the lattice, independent of the total system size. When the moments are known symbolically as functions of the Hamiltonian parameters, the bounds follow by substitution, which makes them a convenient quick benchmark.

\subsection{Recursion method}
\label{sec:recursion}

In practice only the first $n_\text{max}$ moments, or equivalently the first
$n_\text{max}$ Lanczos coefficients, are available. The bounds (\ref{bounds}) provide exact short-time information, but not the full time dependence of $C(t)$. The
recursion method extends this information by completing the Lanczos sequence
with a physically motivated tail. Originally introduced decades ago \cite{Mori_1965_Continued-fraction,Dupuis_1967_Moment,haydock1980recursive,mattis1981reduce,viswanath1994recursion}, the recursion method has recently been used for a number of systems including qubit \cite{uskov2024quantum,shirokov2025quench} and fermionic \cite{ermakov2025symbolic} systems.

In this work we use the recursion method in the following form:
\begin{enumerate}
    \item Compute the first $n_\text{max}$ Lanczos coefficients $b_n$ exactly.
    \item Choose an asymptotic form $b_n^\text{fit}$ consistent with the computed coefficients and with the expected large-$n$ behavior.
    \item Use $b_n^\text{fit}$ to extrapolate the Lanczos sequence up to a cutoff $K$. The cutoff is increased until the resulting $C(t)$ no longer changes on the time interval of interest. We find that $K=10^4$ is sufficient for the time intervals shown.
    \item Substitute the completed sequence into Eq.~(\ref{diffEqCt}) and solve the resulting finite system of differential equations for $\varphi_n(t)$, with $C(t)=\varphi_0(t)$.
\end{enumerate}

Step 2 represents the main source of uncertainty of the method. Indeed, depending on the choice of fit the late-time dynamics of $C(t)$ may vary drastically. The
Universal Operator Growth Hypothesis \cite{parker2019universal} predicts
asymptotically linear growth of $b_n$ in generic chaotic systems, with possible
logarithmic corrections in one dimension. Integrable interacting systems often
show slower growth, for example $b_n\sim\sqrt{n}$. In regimes where independent benchmarks are available, the recursion method has been tested against exact one-dimensional results \cite{uskov2024quantum} and against Majorana-propagation calculations in two dimensions \cite{ermakov2025symbolic}.

\subsection{Operator representation}

In operator-based methods such as the recursion method or Pauli propagation, the main technical task is to manipulate operators in high-dimensional operator spaces. One must store operators efficiently and perform algebraic operations, such as rotations, commutators, and products, as well as nonlinear operations such as norm evaluation.

An operator is represented as
\begin{align}
    \label{operGeneral}
    A=\sum\limits_{k} c_k B_k,
\end{align}
where the $B_k$ are basis elements and the $c_k$ are coefficients. The choice of basis determines both the algebraic cost of the calculation and the cost of evaluating scalar products. The coefficients may be stored in floating-point arithmetic, exact arbitrary-precision arithmetic, or symbolic form. In memory we keep only the nonzero terms, usually as a sparse list or hash map of pairs $\{c_k,B_k\}$. This representation is efficient when the operator remains sparse in the chosen basis, which is precisely the regime targeted by the symbolic and operator-propagation methods used in this work.

\section{Bases of the spin operator space}
\label{sec:basis}

We now specialize to spin systems. We work on a lattice of $L$ sites, each carrying a spin-$S$ degree of freedom, so that the local Hilbert space has dimension $d=2S+1$. The on-site spin operators $S^\alpha_i$ ($\alpha=x,y,z$) obey $[S^\alpha_i,S^\beta_j]=i\delta_{ij}\epsilon_{\alpha\beta\gamma}S^\gamma_i$ (we set $\hbar=1$). We work with finite lattices for convenience, but all results reported in this paper correspond to the thermodynamic limit.

To study operator growth we must fix a basis of the operator space. We use two: the spin monomials, whose algebra is the same for every $S$ and lets us keep the spin magnitude symbolic, and the clock-shift basis, defined at a fixed $S$, orthogonal and closed under multiplication.

\subsection{Basis of spin monomials (non-orthogonal)}

On a single site $i$, a spin monomial is an ordered product of powers of the three spin operators,
\begin{align}
    \label{spinMono}
    m^{abc}_i=(S^x_i)^a(S^y_i)^b(S^z_i)^c,
\end{align}
with non-negative integers $a,b,c$. A basis element on the full lattice is a product of such single-site monomials over all sites,
\begin{align}
    \label{largeM}
    M_{\mathbf{a}\mathbf{b}\mathbf{c}}=\prod_{i=1}^{L} m^{a_i b_i c_i}_i=\prod_{i=1}^{L}(S^x_i)^{a_i}(S^y_i)^{b_i}(S^z_i)^{c_i},
\end{align}
labelled by the exponent vectors $\mathbf{a}=(a_1,\dots,a_L)$, $\mathbf{b}$, $\mathbf{c}$, and an arbitrary operator is a linear combination of them. To make this labelling unique we fix the order of the factors, sites in increasing order $i_1<i_2<\dots$ and $x$ before $y$ before $z$ within each site.

The defining feature of this basis is that its algebra $[S^\alpha,S^\beta]=i\epsilon_{\alpha\beta\gamma}S^\gamma$ does not depend on $S$. We can therefore keep $S$ as a symbolic parameter: products and commutators of monomials are computed once, by reordering the factors into the canonical form (\ref{largeM}) with these relations, and the result holds for every spin at once.

The spin magnitude enters only through the scalar product. The overlap of two monomials (\ref{largeM}) is a polynomial in $x=S(S+1)$ with rational coefficients (Appendix~\ref{app:traces}), so the scalar product of any two operators is again such a polynomial. Because the basis is non-orthogonal, the norm of an operator with $K$ terms is a double sum over all pairs, costing $O(K^2)$, against the $O(K)$ of an orthogonal basis where only identical strings overlap.

The non-orthogonal basis of monomials (\ref{largeM}) lets us compute operators (\ref{operGeneral}) and moments (\ref{moms}) symbolically in $S$, giving results for arbitrary spin. The price is a norm that is costly to evaluate.

\subsection{Clock-shift basis (orthogonal)}

Let us now consider an orthogonal basis for spin systems.

In the context of spin systems the best-known such basis is the many-body basis of Pauli strings at $S=1/2$. A Pauli string is a product of single-site Pauli matrices $\sigma^\alpha_i$, up to a coefficient, and the $4^L$ strings form an orthogonal basis of the operator space. Most operator-based methods use this basis \cite{uskov2024quantum,ermakov2025polynomially,ermakov2024unified,rudolph2026pauli,angrisani2026simulating,schuster2025polynomial}, and efficient software exists to manipulate it \cite{Loizeau2025,lychkovskiy2026qcommute}.

For higher spins $S>1/2$ one can adopt a generalization of the Pauli strings, built from tensor products of shift and clock operators. For a lattice of $L$ qudits with local dimension $d$, a generalized Pauli string is defined as
\begin{align}
    \label{genPaul}
    P(v,w)=X^{v_1}Z^{w_1}\otimes X^{v_2}Z^{w_2}\otimes\dots\otimes X^{v_L}Z^{w_L},
\end{align}
where $v=(v_1,\dots,v_L)$, $w=(w_1,\dots,w_L)$ are integer vectors with entries from $0$ to $d-1$, and $X$, $Z$ are the local shift and clock operators:
\begin{align}
    \label{shiftclock}
    X=\sum\limits^{d-1}_{j=0}|j+1 \bmod d\rangle\langle j|, \qquad Z=\sum\limits^{d-1}_{j=0}\omega^j|j\rangle\langle j|,
\end{align}
with $\omega=e^{2\pi i/d}$. The shift operator $X$ cyclically raises the local state and $Z$ assigns to it a root-of-unity phase. For $d=2$ they reduce to the ordinary Pauli matrices $\sigma_x$ and $\sigma_z$, so that (\ref{genPaul}) is a direct generalization of the qubit Pauli strings. These operators have a long history~\cite{weyl1927quantenmechanik,weyl1950theory,santhanam1976quantum,jagannathan1981finite,jagannathan1982finite,sylvester1909collected} and numerous applications~\cite{gottesman1997stabilizer,gottesman1998fault,gheorghiu2014standard,friedrich2022toolkit,lychkovskiy2021closed,sarkar2024qudit,magni2025quantum}. The full set of operators (\ref{genPaul}) spans the entire operator space of dimension $d^{2L}$ and forms an orthogonal basis under the scalar product (\ref{scalProd}).

A key practical property of this basis, analogous to the standard Pauli strings for qubits, is that the product of any two generalized Pauli strings is again a string, up to a phase:
\begin{align}
    \label{prodPaul}
    P(v,w)P(v',w')=\omega^{\xi(w,v')}P(v'',w''),
\end{align}
where $\xi(w,v')=\sum^L_{i=1}w_iv'_i \bmod d$, and $v''_i=v_i+v'_i \bmod d$, likewise for $w''$. The phase factor arises from the relation $XZ=\omega^{-1}ZX$.

The two orderings $PP'$ and $P'P$ give the same string $P(v'',w'')$ and differ only by their phase, so the commutator of two strings is again a single string,
\begin{align}
    \label{commPaul}
    [P(v,w),P(v',w')]=\big(\omega^{\xi(w,v')}-\omega^{\xi(w',v)}\big)\,P(v'',w'').
\end{align}
This is what makes the basis efficient for operator growth: a commutator with the Hamiltonian never leaves the basis, so nested commutators are evaluated string by string. Being orthogonal, the basis also keeps norms cheap, at a cost linear in the number of terms.

\section{Spin-$S$ Ising model}
\label{sec:ising}

Let us now consider a system described by the Hamiltonian
\begin{align}
    \label{ham}
    H=J\sum\limits_{\langle i,j\rangle}S^x_iS^x_{j}+h_x\sum\limits_{i}S^x_i+h_z\sum\limits_{i}S^z_i,
\end{align}
where $\langle i,j\rangle$ runs over nearest-neighbor pairs of lattice sites and $S^\alpha_i$ ($\alpha=x,y,z$) are spin operators with local Hilbert space dimension $d=2S+1$, satisfying the commutation relations $[S^\alpha_i,S^\beta_j]=i\delta_{ij}\epsilon_{\alpha\beta\gamma}S^\gamma_i$ (we set $\hbar=1$). The parameters $J$, $h_x$ and $h_z$ denote the interaction strength, the transverse field and the longitudinal field. We consider hypercubic lattices in one, two and three dimensions.

As the observable of interest, here and throughout the paper, we consider the total magnetization
\begin{align}
    \label{totmag}
    M=\sum\limits_{i}S^z_i.
\end{align}
We compute the moments (\ref{moms}) as symbolic functions
\begin{align}
    \label{momsym}
    \mu_{2m}(S,J,h_x,h_z)=\frac{\lVert [H(S,J,h_x,h_z),M]^{(m)}\rVert^2}{\lVert M\rVert^2}.
\end{align}
We compute these moments symbolically in the non-orthogonal spin-monomial basis
(\ref{largeM}). The computation is performed directly in the thermodynamic
limit, because it is sufficient to build the nested commutators of a single
local observable $S^z_i$. The moments of $M$ are then recovered from translational invariance, see Appendix \ref{app:thermlim} for details.

\subsection{One-dimensional case}

\begin{figure}[t]
    \centering
    \includegraphics[width=\columnwidth]{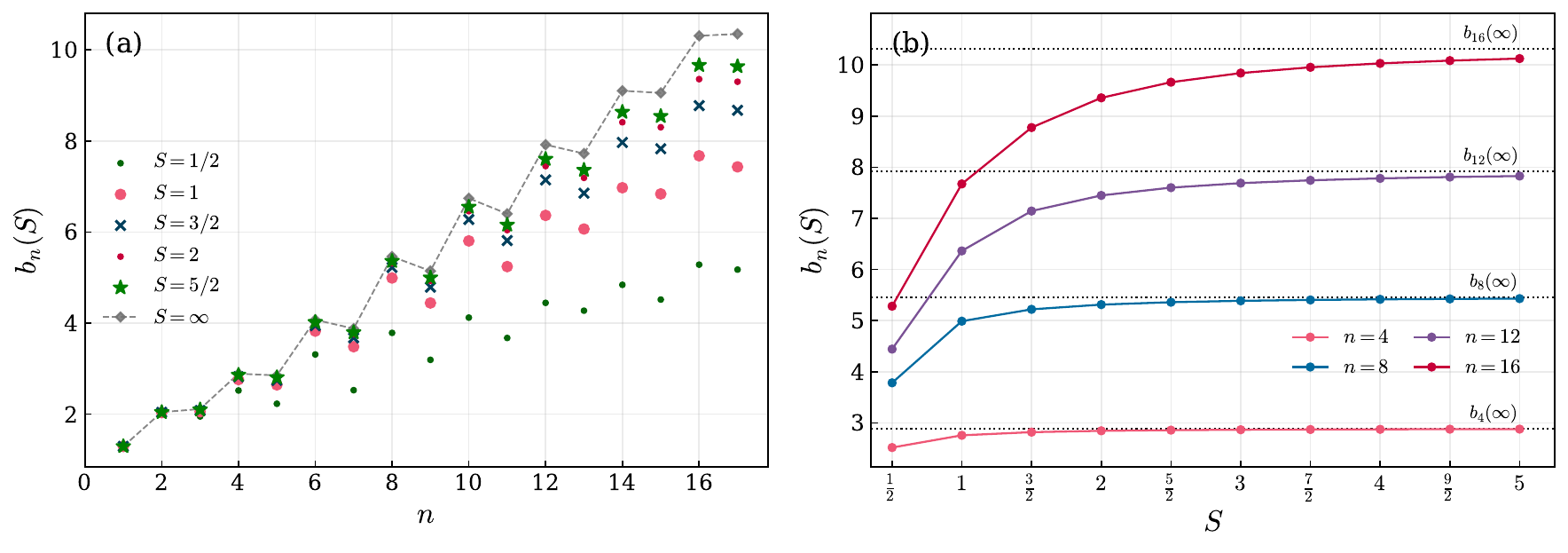}
    \caption{
    Lanczos coefficients $b_n(S)$ for the total magnetization (\ref{totmag}) in the 1D Ising model (\ref{ham}), with $J=1/\sqrt{S(S+1)}$ and $h_x=h_z=1$. (a) $b_n$ as a function of $n$ for $S=1/2,\,1,\,3/2,\,2,\,5/2$, together with the classical limit $b_n(\infty)$ (grey diamonds). (b) $b_n(S)$ as a function of $S$ at fixed $n=4,\,8,\,12,\,16$. Dotted lines mark the $S\to\infty$ asymptotes $b_n(\infty)$.}
    \label{fig:bn_ising}
\end{figure}

For the 1D case of the Hamiltonian (\ref{ham}) we compute symbolically $n_\text{max}=17$ moments. The first three read
\begin{align}
    \label{mom1d}
    \mu_2&=\tfrac{2}{3}J^2x+h_x^2,\notag\\[2pt]
    \mu_4&=\tfrac{16}{15}J^4x^2
          +\left(-\tfrac{2}{15}J^4+4J^2h_x^2+\tfrac{8}{3}J^2h_z^2\right)x
          +h_x^2\left(h_x^2+h_z^2\right),\notag\\[2pt]
    \mu_6&=\tfrac{16}{7}J^6x^3
          +\left(-\tfrac{20}{21}J^6+16J^4h_x^2+\tfrac{128}{9}J^4h_z^2\right)x^2\notag\\
         &\quad+\left(\tfrac{2}{21}J^6-2J^4h_x^2-\tfrac{8}{3}J^4h_z^2
           +10J^2h_x^4+\tfrac{92}{3}J^2h_x^2h_z^2+\tfrac{32}{3}J^2h_z^4\right)x\notag\\
         &\quad+h_x^2\left(h_x^2+h_z^2\right)^2,
\end{align}
where $x=S(S+1)$. Every moment is a polynomial in $x$ with rational coefficients, homogeneous of degree $2m$ in $(J,h_x,h_z)$, and of degree $m$ in $x$. The complete list of moments is available in a public repository~\cite{symboliclab}.

This symbolic computation covers all $S$ at once. For example, setting $x=3/4$ and $\sigma^\alpha=2S^\alpha$ (equivalently $J=4$ and $h_\alpha\to 2h_\alpha$) reduces~(\ref{mom1d}) to the spin-$1/2$ moments of Ref.~\cite{uskov2024quantum} term by term.

From these moments the Lanczos coefficients $b_n$ follow by the standard procedure~\cite{Dupuis_1967_Moment,viswanath1994recursion}. To keep the interaction and the fields on the same scale as $S$ grows, we set $J=1/\sqrt{S(S+1)}$ and $h_x=h_z=1$. Figure~\ref{fig:bn_ising}(a) shows the resulting $b_n$ for $S=1/2,\,1,\,3/2,\,2,\,5/2$ and for the classical limit $S\to\infty$, obtained as the $x\to\infty$ limit of the moments (Sec.~\ref{sec:classicalization}). For every $S$ the coefficients grow linearly with $n$, as predicted by the UOGH.

In one dimension the linear growth is expected to carry a logarithmic correction, seen for this model in Ref.~\cite{uskov2024quantum} at $S=1/2$ and large $n\sim 30$--$40$. Being a feature of the quantum dynamics, this correction should weaken as $S$ grows toward the classical limit, so our $n_\text{max}=17$ is not enough to tell whether it is present at higher spins.

\subsection{Classicalization of Lanczos coefficients}
\label{sec:classicalization}

For the interaction strength $J=1/\sqrt{S(S+1)}=1/\sqrt{x}$ the moments~(\ref{mom1d}) have a classical limit. Each term of $\mu_{2m}$ has the form $c\,J^{2j}h_x^p h_z^q\,x^k$ with $k\le j$, so at $J=1/\sqrt{x}$ it scales as $x^{k-j}$ with $k-j\le 0$. No term grows with $x$, and every moment has a finite limit as $x\to\infty$, therefore
\begin{align}
    \label{momcl}
    \mu_{2m}\big|_{J=1/\sqrt{x}}=\mu^{\mathrm{cl}}_{2m}(h_x,h_z)+O(1/x),
\end{align}
where $\mu^{\mathrm{cl}}_{2m}$ are the classical moments. Writing $S_i=\sqrt{x}\,s_i$ turns the spins into unit vectors with $[s^a_i,s^b_i]=\tfrac{i}{\sqrt{x}}\epsilon_{abc}s^c_i$. With $J=1/\sqrt{x}$ the Hamiltonian becomes $H=\sqrt{x}\,H_{\mathrm{cl}}$ with $H_{\mathrm{cl}}=\sum_i\big(s^x_is^x_{i+1}+h_x s^x_i+h_z s^z_i\big)$, so the $\sqrt{x}$ cancels the $1/\sqrt{x}$ of the commutator and $i[H,\cdot]$ becomes the Poisson bracket as $x\to\infty$. In the same limit the normalized infinite-temperature trace becomes the average over vectors $s_i$ drawn independently and uniformly on the unit sphere. The $\mu^{\mathrm{cl}}_{2m}$ are therefore the moments of the classical autocorrelation function $C_{\mathrm{cl}}(t)=\langle m^z(t)\,m^z(0)\rangle/\langle (m^z)^2\rangle$ of the classical magnetization $m^z=\sum_i s^z_i$, with the spins precessing as $\dot{s}_i=B_i\times s_i$, $B_i=\partial H_{\mathrm{cl}}/\partial s_i$~\cite{Elsayed_2014_Signatures}.

Each $b_n$ is a continuous function of the first $n$ moments, so it inherits this limit,
\begin{align}
    \label{bnclass}
    b_n(S)=b_n(\infty)+O(1/x).
\end{align}
The classical values $b_n(\infty)$ are shown as grey diamonds in Fig.~\ref{fig:bn_ising}(a). Figure~\ref{fig:bn_ising}(b) shows the approach at fixed $n$: the coefficients settle on their classical values already at moderate $S$.

The convergence (\ref{bnclass}) is pointwise in $n$. At finite $S$ a single site carries a finite operator space of $d^2-1$ elements, so the growth in depth is exhausted after a finite number of steps and any further growth has to be spatial. In the classical limit this restriction is absent. The data show the same picture: in Fig.~\ref{fig:bn_ising}(b) the coefficient at $n=4$ reaches its classical value already at $S=1$, while at $n=16$ it still lies below $b_{16}(\infty)$ at $S=5$, so the spin required for classicalization grows with $n$. The limits $n\to\infty$ and $S\to\infty$ therefore do not commute, and the quantum and the classical sequences must eventually differ in their large-$n$ asymptotics.

\subsection{2D and 3D cases}

\begin{figure*}[t]
    \centering
    \includegraphics[width=\textwidth]{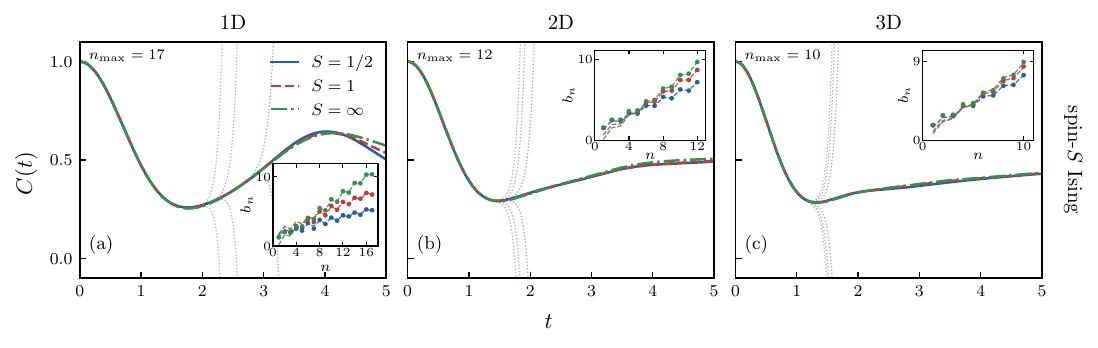}
    \caption{Infinite-temperature autocorrelation function $C(t)$ of the total magnetization~(\ref{totmag}) in the spin-$S$ Ising model~(\ref{ham}), computed by the recursion method of Sec.~\ref{sec:recursion} from the Lanczos coefficients $b_n$ with the tail extrapolated by a fit. Panels: one, two and three dimensions. Solid lines show $C(t)$ from the recursion method, dotted lines the two consecutive truncated-Taylor partial sums that bound $C(t)$ at short times. Insets show the computed $b_n$ (dots) and the fit (dashed), with $n_{\max}$ the number of moments used. Here $h_x=h_z=1$ and $J=1/\sqrt{S(S+1)}$, with $S=1/2$, $1$ and the classical limit $S\to\infty$, and $b_n$ is fitted by $b_n\simeq\alpha n+\gamma+(-1)^n\gamma_\ast$.}
    \label{fig:corrf_ising}
\end{figure*}

The same symbolic procedure can be applied in two and three dimensions. The
computational cost grows rapidly with dimensionality, but the locality of the
Hamiltonian~(\ref{ham}) still makes the thermodynamic-limit calculation
possible: the $m$-th nested commutator of a local operator involves only a
finite region whose size is set by $m$. The resulting moments are available in
the same public repository~\cite{symboliclab}.

The insets of Fig.~\ref{fig:corrf_ising} show the corresponding Lanczos coefficients
at $h_x=h_z=1$ and $J=1/\sqrt{S(S+1)}$. As in one dimension, the coefficients
are consistent with linear growth in $n$. We also observe the same rapid
approach to the classical sequence as $S$ increases.

\subsection{Recursion method and Taylor bounds}
\label{sec:recursion_results}

We now use the recursion method described in Sec.~\ref{sec:recursion} to
reconstruct the infinite-temperature autocorrelation function $C(t)$ from the
Lanczos coefficients. For the spin-$S$ Ising model we fit the computed
coefficients by
\begin{align}
    b_n^\mathrm{fit}=\alpha n+\gamma+(-1)^n\gamma_\ast,
\end{align}
where the last term captures the even-odd oscillations visible at accessible
$n$. We then extrapolate the sequence up to a large cutoff $K$ and solve
Eq.~(\ref{diffEqCt}); in practice we use $K=10000$, which is sufficient for the
time intervals shown.

Figure~\ref{fig:corrf_ising} shows the resulting autocorrelation
functions in one, two and three dimensions. The dotted curves show the
rigorous short-time bounds from Eq.~(\ref{bounds}), obtained from consecutive
Taylor truncations. The recursion-method curves agree with these bounds in
their common validity window and extend the estimate of $C(t)$ to later times.

The same classicalization pattern observed in the Lanczos coefficients is also
visible in the correlation functions. At early and intermediate times the
finite-$S$ curves are already close to the classical limit. At later times the curves begin to separate. This later-time regime is more sensitive to the large-$n$ part of the Lanczos sequence, where the finite-$S$ data and the classical limit are no longer as close as at small $n$.

\section{$q$-state Potts model}
\label{sec:potts}

The spin-$S$ Ising model has a natural classical limit, which can be accessed within the spin-monomial basis. We now consider a different qudit model, the $q$-state quantum Potts model. Unlike the spin-$S$ Ising model, it is formulated in terms of generalized Pauli strings (\ref{genPaul}) and has no $S\to\infty$ classical limit. The Hamiltonian of the $q$-state Potts model reads
\begin{align}
H_{\mathrm{Potts}}
=
-J
\sum_{\langle i,j\rangle}
\sum_{k=1}^{q-1}
\left(
Z_{i}Z_{j}^{\dagger}
\right)^k
-
h
\sum_{i}
\sum_{k=1}^{q-1}
X_{i}^{k},
\label{ham:quantum_potts_compact}
\end{align}
where $X$ and $Z$ are the shift and clock operators defined in
Eq.~(\ref{shiftclock}). In this section we denote the local Hilbert-space dimension by $q$, following the standard terminology of the $q$-state Potts model, which is the same dimension denoted by $d$ in Sec.~\ref{sec:basis}.

As the observable we again use the total magnetization (\ref{totmag}), which in the basis (\ref{genPaul}) takes the form
\begin{align}
    M=\sum_i\sum_{k=1}^{q-1}\frac{1}{1-\omega^{-k}}\,Z_i^k .
    \label{totmag_clock}
\end{align}
For $q=2$ each site contributes $S^z_i=Z_i/2$, as expected.

As in the previous case, we compute the moments of the infinite-temperature autocorrelation function symbolically as functions of the Hamiltonian parameters. Here the local dimension $q$ is fixed, while the dependence on $J$ and $h$ is kept symbolic. The calculation is performed in the orthogonal basis (\ref{genPaul}). All the results are valid in the thermodynamic limit, see Appendix~\ref{app:thermlim} for details.

\subsection{One-dimensional case}

\begin{figure}[t]
    \centering
    \includegraphics[width=\columnwidth]{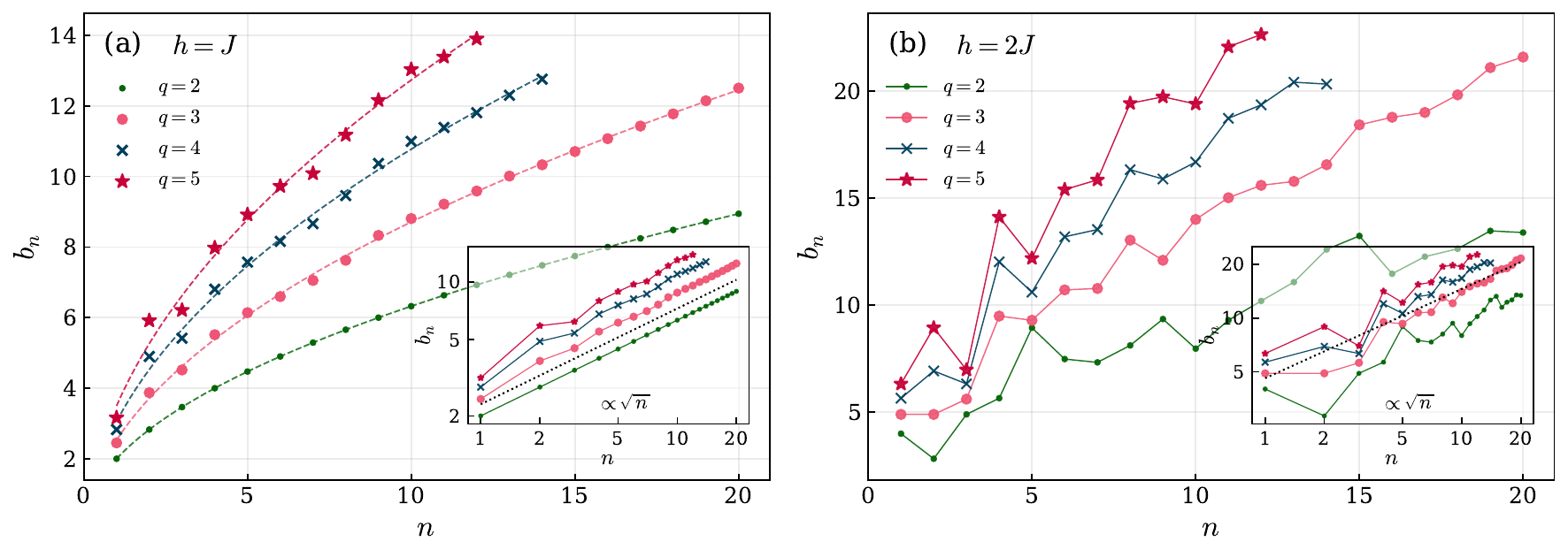}
    \caption{
    Lanczos coefficients $b_n$ for the total magnetization (\ref{totmag_clock}) in the 1D $q$-state Potts model (\ref{ham:quantum_potts_compact}), for $q=2,\,3,\,4,\,5$. (a) Self-dual point $h=J$, with the fit $b^\mathrm{fit}_n=\alpha+\gamma\sqrt{n}$ shown as dashed lines and $(\alpha,\gamma)=(0,2)$, $(-0.33,2.86)$, $(-0.52,3.57)$, $(-0.78,4.27)$ for $q=2,\,3,\,4,\,5$. At $q=2$ the fit is exact $b_n=2\sqrt{n}$. (b) $h=2J$. Insets: the same coefficients on logarithmic axes. Dotted reference line of slope $1/2$. The coefficients are computed up to the $n_{\max}$ of Table~\ref{tab:nmax_potts}.}
    \label{fig:bn_potts}
\end{figure}

In one dimension the first three moments read, for several values of $q$,
\begin{subequations}
\label{mom_potts_1d}
\begin{align}
&\mu_2^{(q=3)} = 6h^2, \notag\\
&\mu_4^{(q=3)} = 72J^2h^2+54h^4, \notag\\
&\mu_6^{(q=3)} = 1944J^4h^2-216J^3h^3
              +2268J^2h^4+486h^6,
\end{align}
\begin{align}
&\mu_2^{(q=4)} = 8h^2, \notag\\
&\mu_4^{(q=4)} = 128J^2h^2+128h^4, \notag\\
&\mu_6^{(q=4)} = 5120J^4h^2-1024J^3h^3
              +7680J^2h^4+2048h^6,
\end{align}
\begin{align}
&\mu_2^{(q=5)} = 10h^2, \notag\\
&\mu_4^{(q=5)} = 200J^2h^2+250h^4, \notag\\
&\mu_6^{(q=5)} = 11000J^4h^2-3000J^3h^3
              +19500J^2h^4+6250h^6.
\end{align}
\end{subequations}

The moments (\ref{mom_potts_1d}) are again polynomials in the Hamiltonian parameters $J$ and $h$, here with integer coefficients. The remaining moments are available in the repository~\cite{symboliclab}. The number of moments $n_{\max}$ we reach depends on the local dimension $q$ and on the dimensionality of the lattice, and is listed in Table~\ref{tab:nmax_potts}.

\begin{table}[t]
\centering
$n_{\max}$ computed\\[3pt]
\begin{tabular}{lcccc}
\hline
       & $q=2$ & $q=3$ & $q=4$ & $q=5$ \\
\hline
1D     & 50 & 20 & 14 & 12 \\
2D     & 19 & 12 & 10 & -- \\
3D     & 14 & 10 & -- & -- \\
\hline
\end{tabular}
\caption{Number of moments $n_{\max}$ computed for the $q$-state Potts model (\ref{ham:quantum_potts_compact}).}
\label{tab:nmax_potts}
\end{table}

Figure~\ref{fig:bn_potts} shows the Lanczos coefficients obtained from these moments. At the self-dual point $h=J$ they are well described by $b^\text{fit}_n=\alpha+\gamma\sqrt{n}$, with small oscillations on top. Away from the self-dual point, for example at $h=2J$, the fit is less accurate, although the inset shows that the asymptotics remains square-root like.

In one dimension the Hamiltonian (\ref{ham:quantum_potts_compact}) is integrable, and the linear asymptotic growth of $b_n$ predicted by the UOGH is therefore not expected. The data of Fig.~\ref{fig:bn_potts} are a clear example of sublinear asymptotics of the Lanczos coefficients. For $q=2$ the model is the transverse-field Ising chain, which is free-fermionic, yet it is known that the growth of $b_n$ for the observable (\ref{totmag_clock}) is the same as in an interacting integrable model \cite{parker2019universal,ermakov2025polynomially}.

\subsection{Two- and three-dimensional cases}

We also compute the symbolic moments of the Hamiltonian (\ref{ham:quantum_potts_compact}) in two and three dimensions, for the same observable (\ref{totmag_clock}).

There the Hamiltonian is no longer integrable, and it is therefore reasonable to expect the linear growth of the Lanczos coefficients predicted by the UOGH. Indeed, the insets of Fig.~\ref{fig:corrf_potts}(b,c) show that $b_n$ follows the linear asymptotics $b_n=\alpha n+\gamma$ to good accuracy. These data confirm the prediction of the UOGH in models with local Hilbert-space dimension larger than two.

\subsection{Recursion method and Taylor bounds}

\begin{figure*}[t]
    \centering
    \includegraphics[width=\textwidth]{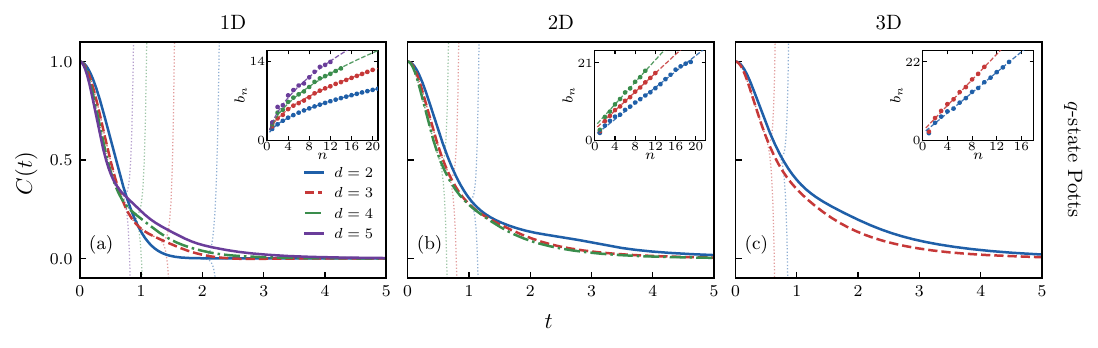}
    \caption{Infinite-temperature autocorrelation function $C(t)$ of the total magnetization~(\ref{totmag_clock}) in the $q$-state Potts model~(\ref{ham:quantum_potts_compact}) at $J=h=1$, computed by the recursion method of Sec.~\ref{sec:recursion} from the Lanczos coefficients $b_n$ with the tail extrapolated by a fit. Panels: one, two and three dimensions. Solid lines show $C(t)$ from the recursion method, dotted lines the two consecutive truncated-Taylor partial sums that bound $C(t)$ at short times. Insets show the computed $b_n$ (dots) and the fit (dashed), the latter continued past the last computed coefficient to display the extrapolation. The fit is $b_n\simeq\alpha+\gamma\sqrt{n}$ in one dimension and $b_n\simeq\alpha n+\gamma$ in two and three dimensions.}
    \label{fig:corrf_potts}
\end{figure*}

Lastly, we again use the symbolic data to compute infinite-temperature autocorrelation functions. Figure~\ref{fig:corrf_potts} shows $C(t)$ obtained from the recursion method, together with the Taylor bounds (\ref{bounds}) on its initial behavior.

In one dimension we extrapolate the Lanczos coefficients with the square-root fit $b^\text{fit}_n=\alpha+\gamma\sqrt{n}$, and in two and three dimensions with a linear one. In every case the correlation function decays to zero, $C(t\rightarrow\infty)=0$. Increasing the local dimension $q$ changes neither the character of the operator growth nor that of the decay of $C(t)$.

\section{Discussion}
\label{sec:discussion}
The symbolic moments obtained here are exact and valid directly in the thermodynamic limit, with no arithmetic error and no extrapolation in the system size. This makes them a natural benchmark. They are also compact, so that anyone can download them \cite{symboliclab} and use them as they are, either for the rigorous Taylor bounds on the initial dynamics or as input to the recursion method. Obtaining the same numbers by other means would require exact diagonalization of a system large enough for the boundary not to be felt, which for $n_{\max}=10$ in three dimensions already means a lattice of about $21\times21\times21$ sites, by no means accessible even for spin-$1/2$ systems.

Our results support the Universal Operator Growth Hypothesis beyond spin-$1/2$ and beyond local dimension two. In the non-integrable cases the Lanczos coefficients grow linearly with $n$, while in the integrable one-dimensional Potts chain the growth is sublinear.

Classicalization here is a property not of a single observable but of the whole Lanczos sequence, that is of the mechanism of operator growth itself. It also explains why classical simulations of infinite-temperature autocorrelation functions work well already at moderate spin.

It is also interesting to think of the logarithmic corrections observed in the asymptotics of $b_n$ in 1D spin models~\cite{parker2019universal,uskov2024quantum}. The nature of these corrections is rather heuristic and is typically associated with the limited room available to a growing operator on a line. We therefore expect them to move to larger $n$ as $S$ grows and to vanish in the classical limit, which further stresses that the large-$n$ and large-$S$ limits cannot be interchanged. The depth reached here is far from sufficient to test this, so we state it as an expectation.

The $n_\text{max}$ reported here, for most of the cases excluding $S=1/2$, is to our knowledge the largest available for these models to date. This depth is relatively high and allows us to analyse the asymptotics. In each specific case, however, we did not push $n_\text{max}$ to the limit, as that would require further code optimization and more computational resources and time. All the results were obtained within 1TB of RAM.

The machinery developed here is not restricted to the recursion method. The same engine can be used for unitary operator propagation and for truncation-based methods in qudit bases. Operator propagation is moreover no longer confined to real-time dynamics: it has recently been applied to ground-state energy estimation in strongly correlated models, competitively with DMRG in two dimensions \cite{shrikhande2026pauli,li2026pauli}. Carrying this over to qudits would put the low-temperature properties of higher-spin models within reach.

\section*{Acknowledgements}
The author is grateful to Oleg Lychkovskiy and Vsevolod Yashin for useful discussions, and especially to Oleg Lychkovskiy for introducing me to the recursion method, which laid the foundation for my work in this area. \\
The work was supported by Rosatom in the framework of the Roadmap for Quantum computing (Contract No. 868-1.3-15/15-2021 dated October 5, 2021). Part of the work present in the paragraph ``q-state Potts model'' was supported by the Foundation for the Advancement of Theoretical Physics and Mathematics ``BASIS'' under the grant N 22-1-2-55-1.

\appendix

\section{Scalar products in the spin-monomial basis}
\label{app:traces}

Here we describe how the scalar product (\ref{scalProd}) of two spin monomials is evaluated exactly, as a polynomial in $x=S(S+1)$.

\subsection{Factorization over sites}

Recall the infinite-temperature scalar product (\ref{scalProd}), $(A|B)=\mathrm{tr}(A^\dagger B)/d^L$ with $d=2S+1$. Any operator $A$ can be expanded in monomials
\begin{align}
    \label{App1:largeM}
    M_{\mathbf{a}\mathbf{b}\mathbf{c}}=\prod_{i=1}^{L} m^{a_i b_i c_i}_i=\prod_{i=1}^{L}(S^x_i)^{a_i}(S^y_i)^{b_i}(S^z_i)^{c_i}.
\end{align}
The scalar product of two monomials factorizes over sites as
\begin{align}
    \label{App1:factorize}
    (M\,|\,M')=\frac{\mathrm{tr}(M^\dagger M')}{d^L}=\prod_{i=1}^{L}\frac{\mathrm{tr}(m_i^\dagger m'_i)}{d},
\end{align}
indeed, the trace of a tensor product is the product of the traces, and $\dim(\mathcal H)=d^L$.

For operators $A=\sum_k c_k M_k$ and $B=\sum_l e_l M_l$ the scalar product is a sum over pairs of terms:
\begin{align}
    \label{App1:innerpairs}
    (A|B)=\sum_{k,l} c^*_k\,e_l\prod_{i=1}^{L} g\big(m^i_k,m^i_l\big),
\end{align}
built from the single-site overlap
\begin{align}
    \label{App1:gsite}
    g(m,m')=\frac{\mathrm{tr}(m^\dagger m')}{d}=\frac{1}{d}\,\mathrm{tr}\big[(S^z)^{c}(S^y)^{b}(S^x)^{a}\,(S^x)^{a'}(S^y)^{b'}(S^z)^{c'}\big].
\end{align}
The argument of the trace in the above formula must be rewritten in normal order $x<y<z$. Applying the commutator $[S^\alpha,S^\beta]=i\epsilon_{\alpha\beta\gamma}S^\gamma$ repeatedly, we obtain
\begin{align}
    \label{normord}
    (S^z)^{c}(S^y)^{b}(S^x)^{a}(S^x)^{a'}(S^y)^{b'}(S^z)^{c'}=\sum_{A,B,C}\gamma_{ABC}\,(S^x)^A(S^y)^B(S^z)^C.
\end{align}
The coefficients $\gamma_{ABC}$ are complex numbers with integer real and imaginary parts (so-called Gaussian integers, $\gamma_{ABC}\in\mathbb{Z}[i]$), and they do not depend on $S$, because the commutator that drives the reordering is itself $S$-independent.

\subsection{Single-monomial traces}

Combining (\ref{App1:gsite}) and (\ref{normord}), the overlap reduces to the computation of single-monomial traces:
\begin{align}
    \label{gasN}
    g(m,m')=\sum_{A,B,C}\gamma_{ABC}\,N_{ABC}(x),
\end{align}
where
\begin{align}
    \label{App1:Nform}
    N_{ABC}(x)\equiv\frac{1}{d}\,\mathrm{tr}\big[(S^x)^A(S^y)^B(S^z)^C\big]
\end{align}
is the single-monomial trace.

Let us prove two simple lemmas about these traces.

\begin{lemma}[Parity]
\label{App1:lem-parity}
$N_{ABC}$ vanishes unless $A,B,C$ are all even or all odd. It is then real if they are all even and purely imaginary if all odd.
\end{lemma}

\begin{proof}
Let us consider a rotation $R_z(\pi)=e^{-i\pi S^z}$ that flips $S^x\to-S^x$, $S^y\to-S^y$ and keeps $S^z$. Such a unitary rotation leaves the trace unchanged, so $N_{ABC}=(-1)^{A+B}N_{ABC}$. Now let us consider the rotation $R_x(\pi)=e^{-i\pi S^x}$ about the $x$ axis, which gives $(-1)^{B+C}$. Both signs can equal $+1$ only if $A,B,C$ share a parity.

In the $S^z$ basis matrices $S^x$ and $S^z$ are real and $S^y$ is imaginary, therefore complex conjugation gives $N_{ABC}^{*}=(-1)^{B}N_{ABC}$: real for $B$ even, imaginary for $B$ odd.
\end{proof}

\begin{lemma}[Polynomiality]
\label{App1:lem-poly}
$N_{ABC}$ is a polynomial in $x=S(S+1)$ with rational, possibly imaginary, coefficients, of degree at most $\lfloor(A+B+C)/2\rfloor$.
\end{lemma}

This property is often quoted as a well-known fact. One might also argue that it is clear from the existence of the Casimir element $\mathcal{C}=(S^x)^2+(S^y)^2+(S^z)^2$, which commutes with every polynomial in the spin operators and equals $x$ in the spin-$S$ representation. A complete proof, however, is not easy to find in the literature. It seems that Lemma~\ref{App1:lem-poly} was first proven in Ref.~\cite{bose1964traces}. Following that work, we prove it here.

\begin{proof}
Let us introduce ladder operators
\begin{align}
    \label{App1:ladder}
    S^\pm=S^x\pm iS^y,
    \qquad
    S^x=\frac{S^++S^-}{2},
    \qquad
    S^y=\frac{S^+-S^-}{2i}.
\end{align}
Substituting this into $(S^x)^A(S^y)^B(S^z)^C$ and expanding the brackets turns the monomial into a sum of products of $S^+$, $S^-$ and $S^z$. The numerical prefactors come from the $1/2$ and $1/2i$ in (\ref{App1:ladder}), so they are rational, up to an overall factor of $i$.

For these operators $S^\pm|m\rangle\propto|m\pm1\rangle$, so a product with unequal numbers of raising and lowering operators shifts $m$ and has no diagonal matrix elements. Its trace is zero, and only products with equally many raising and lowering operators need to be kept.

Two identities let us remove the ladder operators. From $[S^z,S^\pm]=\pm S^\pm$,
\begin{align}
    \label{App1:shift}
    (S^z)^n\,S^\pm=S^\pm\,(S^z\pm1)^n,
\end{align}
so a power of $S^z$ can be moved through a ladder operator, shifting $S^z\to S^z\pm1$. Multiplying the two ladder operators against each other and using $(S^x)^2+(S^y)^2+(S^z)^2=x$,
\begin{align}
    \label{App1:pair}
    S^+S^-=x-(S^z)^2+S^z,
    \qquad
    S^-S^+=x-(S^z)^2-S^z.
\end{align}
An adjacent raising-lowering pair thus becomes an ordinary polynomial in $S^z$, at the cost of one factor of $x$.

Now move all powers of $S^z$ to the right using (\ref{App1:shift}). What remains on the left is a string of $S^+$ and $S^-$ alone, and since both are present, some $S^+$ stands next to an $S^-$. Identity (\ref{App1:pair}) turns that pair into a polynomial in $S^z$, which we move to the right again. Each round removes one $S^+$ and one $S^-$, so after finitely many rounds no ladder operator is left and
\begin{align}
    \label{App1:reduction}
    (S^x)^A(S^y)^B(S^z)^C=\sum_{r,j}c_{rj}\,x^{r}(S^z)^{j}+(\text{terms with unequal }S^+,S^-),
\end{align}
with rational coefficients $c_{rj}$. The terms with unequal numbers of $S^+$ and $S^-$ vanish under the trace. Counting $x$ as two operators and $S^z$ as one, neither (\ref{App1:shift}) nor (\ref{App1:pair}) raises the count, so every term keeps $2r+j\le k$ with $k=A+B+C$.

Taking the trace of (\ref{App1:reduction}), the unbalanced terms vanish and
\begin{align}
    \label{App1:Ntrace}
    N_{ABC}=\sum_{r,j}c_{rj}\,x^{r}p_j,
    \qquad
    p_j\equiv\frac1d\,\mathrm{tr}\,(S^z)^j=\frac{1}{2S+1}\sum_{m=-S}^{S}m^{j}.
\end{align}
Since $S^z$ is diagonal with entries $m=-S,\dots,S$, each $p_j$ is a rational number. It vanishes for odd $j$ by the symmetry $m\to-m$, and for even $j$ it is a polynomial in $x$ of degree $j/2$, a standard property of these power sums, for instance $p_2=x/3$ and $p_4=x(3x-1)/15$. Together with $2r+j\le k$ this makes $N_{ABC}$ a polynomial in $x$ of degree at most $\lfloor k/2\rfloor$ with rational coefficients. By Lemma~\ref{App1:lem-parity} it is real when $A,B,C$ are all even and imaginary when they are all odd.
\end{proof}

Since the $\gamma_{ABC}$ are $S$-independent and each $N_{ABC}$ is a polynomial in $x$, the single-site overlap $g(m,m')$, and hence every moment, is a polynomial in $x$.

\subsection{Exact evaluation of $N_{ABC}$}

By Lemma~\ref{App1:lem-poly} each single-monomial trace is a polynomial in $x$ of known degree,
\begin{align}
    \label{App1:Npoly}
    N_{ABC}(x)=\sum_{i=0}^{D}c_i\,x^i,\qquad D=\left\lfloor\frac{A+B+C}{2}\right\rfloor,
\end{align}
with rational coefficients $c_i$. It remains only to find these coefficients. For a fixed spin $S$ the left-hand side can be computed directly: we build the $(2S+1)$-dimensional matrices $S^x,S^y,S^z$, form the product, and take its normalized trace, all in high-precision floating point (we use a $1024$-bit mantissa, about $300$ decimal digits). Evaluating at $S=\tfrac12,1,\tfrac32,\dots$ gives $N_{ABC}$ at as many distinct points $x=S(S+1)$ as we need, and the $c_i$ follow from a Vandermonde solve.

Although these values are computed numerically, they are recovered exactly. Each sampled value of $N_{ABC}$ is rational, purely imaginary with a rational modulus when $A,B,C$ are all odd, with a small denominator bounded by those of the power sums $p_j$ that enter the reduction. Continued-fraction reconstruction returns such a rational $p/q$ from a numerical value once the error drops below $1/(2q^2)$, a bound that $300$ digits clear by an enormous margin, so each sampled value, and hence each coefficient $c_i$, is obtained exactly. As an independent check we compute the trace at one extra spin and require the polynomial (\ref{App1:Npoly}) to reproduce it. A wrong reconstruction would fail this test. In practice it always passes, and no floating-point error survives into the coefficients that enter the moments.

\subsection{Computing the norm}

We store an operator as a list of its terms, each with a coefficient. On a single site a monomial $(S^x)^a(S^y)^b(S^z)^c$ is recorded by its three powers $(a,b,c)$, and a term is recorded by the powers on each site it occupies, with identity sites left out. Writing $(a,b,c)_i$ for the powers on site $i$, the operator $A=\sum_k c_k M_k$ becomes the nested list
\begin{align}
    \label{App1:layout}
    A\;\longleftrightarrow\;\Big\{\ \big\{\,c_1,\,(a,b,c)_{i_1},\,(a,b,c)_{i_2},\dots\big\},\ \big\{\,c_2,\,\dots\big\},\ \dots\ \Big\},
\end{align}
one inner list per term, holding its coefficient and the factors on the sites it occupies, ordered by site.

The norm is the double sum (\ref{App1:innerpairs}). We run over pairs of terms $(k,l)$ and multiply the single-site overlaps $g(m^i_k,m^i_l)$ of (\ref{App1:gsite}) across the sites. By the parity rule of Lemma~\ref{App1:lem-parity} a pair contributes only when the two monomials share a parity class at every site, which discards most of the pairs. Each overlap is evaluated by bringing the product $m^\dagger m'$ to canonical order with the commutation relations, as in (\ref{normord}), and weighting the resulting monomials by the traces $N_{ABC}$, which gives (\ref{gasN}). This completes the evaluation of the scalar product.

\section{Details of calculations in the thermodynamic limit}
\label{app:thermlim}

The total-magnetization operator $M=\sum_i S^z_i$ of Eq.~(\ref{totmag}) is a sum
over the whole lattice. Here we explain how one can obtain $\lVert [H,M]^{(n)}\rVert^2$ per lattice site in the thermodynamic limit.

\subsection{Locality fixes the lattice size}

We consider Hamiltonians with nearest-neighbor coupling. If an operator is
supported on a set of sites $R$, then $[H,A]$ is supported on $R$
together with the sites adjacent to it. Each commutator extends the support by one lattice spacing, so an operator seeded on a single site stays within distance $n$ of that site after $n$ commutators, occupying at most $2n+1$ sites along each lattice direction. This holds in any dimension.

It is therefore sufficient to compute the nested commutators $[H,A]^{(n)}$ on a lattice with $L>4n_\text{max}$ and to place the seed at its centre. The operator
$[H,A]^{(n)}$ never reaches the boundary, and every coefficient produced equals the one the same calculation gives on the infinite lattice.

\subsection{From a local seed to the extensive sum}

Let $T_r$ denote the translation by $r$ sites, so that the observable
(\ref{totmag}) is the sum of the translates of a single-site operator,
$M=\sum_r T_r S^z_0$. Write
\begin{align}
    \label{app2:seed}
    O_n\equiv[H,S^z_0]^{(n)}
\end{align}
for the $n$-fold commutator of that one seed.

The Hamiltonian is translation invariant, so $[H,\cdot]$ commutes with $T_r$ and
\begin{align}
    \label{app2:transl}
    [H,M]^{(n)}=\sum_r T_r O_n .
\end{align}
Only $O_n$ has to be computed, and it has finite support.

The translation invariance of the scalar product (\ref{scalProd}) then gives
\begin{align}
    \label{app2:norm}
    \lVert[H,M]^{(n)}\rVert^2=\sum_{r,r'}(T_rO_n|T_{r'}O_n)=N\sum_r (O_n|T_rO_n),
\end{align}
where $N$ is the number of sites. It cancels in the moment (\ref{moms}),
\begin{align}
    \label{app2:moment}
    \mu_{2n}=\frac{\lVert[H,M]^{(n)}\rVert^2}{\lVert M\rVert^2}
    =\frac{\sum_r (O_n|T_rO_n)}{\sum_r (S^z_0|T_rS^z_0)} .
\end{align}
Both sums are finite: $O_n$ has finite support, so $(O_n|T_rO_n)$ vanishes as soon
as $r$ moves the two copies apart. Neither side refers to the lattice size.

\subsection{Folding}

The sum over translations in (\ref{app2:moment}) is evaluated by folding $O_n$.
Each term is translated on its own so that its first occupied site sits at a fixed
reference position, and terms that become equal are added.

Two terms that differ by a translation are thereby mapped onto the same string,
\begin{align}
    \label{app2:foldex}
    c_1\,X_3Z_4\;\longmapsto\;c_1\,X_0Z_1,
    \qquad
    c_2\,X_7Z_8\;\longmapsto\;c_2\,X_0Z_1,
\end{align}
and enter the folded operator as $(c_1+c_2)\,X_0Z_1$, while terms that are not
translates of one another keep separate representatives. Denoting the folded
operator by $\bar O_n$,
\begin{align}
    \label{app2:fold}
    \lVert\bar O_n\rVert^2=\sum_r (O_n|T_rO_n).
\end{align}
This holds because the basis is orthogonal: on either side only terms of the same
translation class contribute, and each class enters with the square of the sum of
its coefficients.

The fold may be applied after every commutation rather than once at the end,
because $[H,\cdot]$ commutes with translations and therefore maps translation
classes to translation classes. The condition $L>4n_\text{max}$ also ensures that
no two terms are identified through the periodic boundary, so the folded operator
is the one of the infinite lattice.

\subsection{Non-orthogonal basis}

In the spin-monomial basis of Sec.~\ref{sec:basis} distinct monomials are not
orthogonal, so $\lVert\bar O_n\rVert^2$ is not the per-site moment and
(\ref{app2:fold}) does not apply. Instead we should consider the sum
\begin{align}
    \label{app2:qzero}
    \widetilde O_n=\sum_{r=-R}^{R}T_r\bar O_n ,
\end{align}
where $R$ is the support width of $\bar O_n$, that is the largest shift for which
the two copies still overlap. The per-site moment is then a single scalar product,
\begin{align}
    \label{app2:nonorth}
    \sum_r (O_n|T_rO_n)=(\bar O_n|\widetilde O_n),
\end{align}
evaluated with the traces of Appendix~\ref{app:traces}.

\bibliography{ref}

\end{document}